\documentclass[12pt,a4paper,dvips]{article}
\usepackage{epsfig}
\usepackage{ a4p}
\usepackage{cite,mcite}
\usepackage{graphicx}
\usepackage{ physics, l3_title,ifthen, Lep}
\journalname{Phys. Lett. B}
\date{November 10, 2000}
\preprint{2000-144}
\newlength{\capindent}
\newlength{\capwidth}
\newlength{\figwidth}
\newcommand{\icaption}[2][!*!,!]{\hspace*{\capindent}%
  \begin{minipage}{\capwidth}
    \ifthenelse{\equal{#1}{!*!,!}}%
      {\caption{#2}}%
      {\caption[#1]{#2}}
  \end{minipage}}

\def\NPB{{Nucl. Phys.} {\bf B }}
\def\PLB{{Phys. Lett.} {\bf B }}

\def\PRD{{Phys. Rev.} {\bf D }}
\def\ZPC{{Z. Phys.} {\bf C }}
\def\CPC{Comp. Phys. Comm. }

\def\ra{\rightarrow}

\def\be{\begin{equation}}
\def\ee{\end{equation}}
\def\bea{\begin{eqnarray}}
\def\eea{\end{eqnarray}}

\begin{document}
\bibliographystyle{l3style}
\begin{titlepage}
\title{Measurements of the Cross Sections for Open Charm and
Beauty Production in \boldmath{${\gamma \gamma}$} Collisions at
$\mathrm{\bf\sqrt{s}=189-202}$ GeV}
\author{The L3 Collaboration}
%
%

\begin{abstract}

The production of c and b quarks in $\mathrm{\gamma\gamma}$ collisions
is studied with the L3 detector at LEP with 410 $\mathrm{pb^{-1}}$ of data,
collected at centre-of-mass energies from 189 GeV to 202 GeV. 
Hadronic final
states containing c and b quarks are identified by detecting electrons
or muons from their semileptonic decays. The cross sections 
$\mathrm{\sigma(e^+e^-\rightarrow e^+e^-c \bar c X)}$ 
and $\mathrm{\sigma(e^+e^-\rightarrow e^+e^-b \bar b X)}$ are 
measured and compared to
next-to-leading order perturbative QCD calculations. The cross
section of b production 
is measured in $\mathrm{\gamma\gamma}$ collisions for the first time.
It is in excess of the QCD prediction by a factor of three.

\end{abstract}
%
%
\submitted

\end{titlepage}
\section{Introduction}

\begin{figure}[htbp]
\begin{center}
 \mbox{\epsfig{file=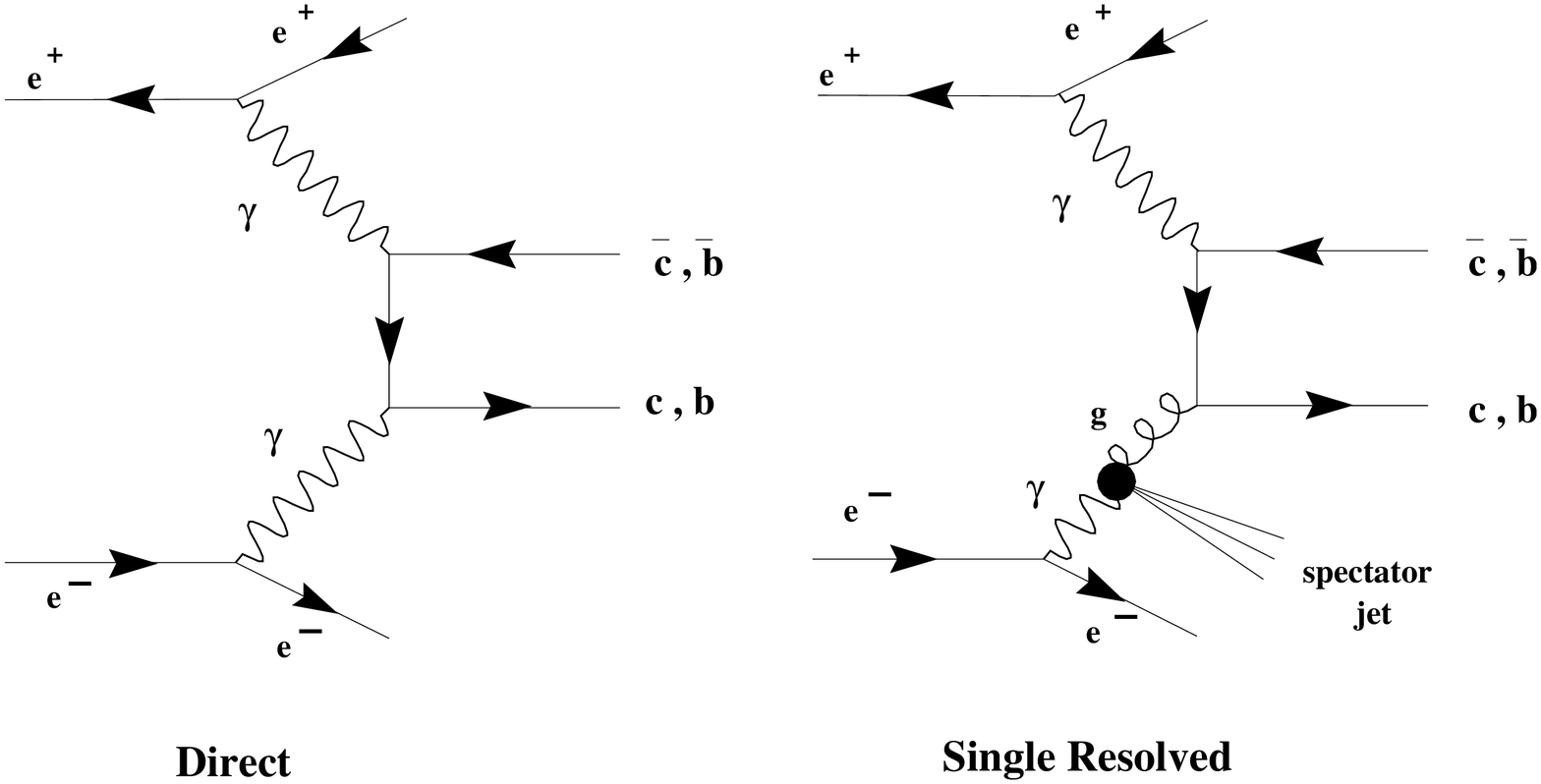, width=.9\textwidth}}  
  \caption{Diagrams contributing to charm and beauty
production in $\mathrm{\gamma \gamma}$
    collisions at LEP.}
  \label{fig:Feynman}
\end{center}
\end{figure}

The measurement of
heavy flavour production in two-photon collisions provides a reliable 
test of
perturbative QCD because of the large physical scale set by the charm or 
beauty quark mass.
Many experiments have studied charm production in $\mathrm{\gamma\gamma}$
collisions~\cite{previous, L3}. Beauty quark production in $\mathrm{\gamma\gamma}$
collisions is expected to be suppressed by more than two orders of magnitude
owing to its smaller electric charge and larger mass.
At LEP
energies, the direct and single resolved processes,
shown in Figure~\ref{fig:Feynman}, are predicted to give comparable
contributions to the heavy
flavour production cross section \cite{theory}, whereas 
at low energies the direct process dominates.
The main contribution to the resolved photon
cross section is the photon-gluon fusion process 
$\mathrm{\gamma g \rightarrow  c \bar{c}(b \bar{b})}$. 
The production rate of c and b quarks in
two-photon collisions therefore depends on their mass and
the gluon density in the photon. Further contributions to 
charm production arise from
the Vector Dominance Model (VDM) and from doubly 
resolved processes and are expected to be small. 
  
This letter describes the measurements of the
inclusive open charm and beauty production
performed with the
L3 detector~\cite{L3det} 
with 410 $\mathrm{pb^{-1}}$ of data,
collected at centre-of-mass energies from 189 GeV to 202 GeV.

We identify b and c quarks by
tagging electrons
\footnote{Electron stands for electron or positron throughout this
paper.} 
or muons from their semileptonic decays.
Due to the higher mass of the b quarks,
these leptons are characterized by 
a higher momentum and higher transverse momentum than those
from c quarks. The $\mathrm{b \bar{b}}$ production
cross section is hence measured 
using the muon and electron transverse momentum spectra 
for both leptons having a momentum greater than 2 GeV. The charm
cross section measurement is made by analysing the electron
spectrum with momentum down to 0.6 GeV.
Muons are always required to have a momentum greater than 2 GeV
necessary to penetrate the calorimeters and reach the
muon chambers. 

\section{Monte Carlo}

The PYTHIA~\cite{pythia} Monte Carlo is
used to model the two photon processes.
The non-b events
are generated with massless matrix
elements~\cite{massless} while for b events
massive matrix elements are used. The resolved process uses the SaS1d photon
structure function~\cite{sas1d}.   
The two-photon luminosity function is implemented in the 
equivalent photon approximation (EPA)~\cite{Budnev} with a 
cutoff $\mathrm{Q^2 < m_{\rho}^2}$ .  

 Background sources are 
$\mathrm{e^{+}e^{-} \ra e^{+}e^{-} \tau^{+} \tau^{-}}$, 
$\mathrm{e^{+}e^{-} \rightarrow Z/\gamma \ra q \bar{q}}$, 
$\mathrm{e^{+}e^{-} \ra \tau^{+} \tau^{-}}$
and $\mathrm{e^{+}e^{-}} \ra$ \newline $\mathrm{W^{+} W^{-}}$.
These processes are generated with JAMVG~\cite{verm}, PYTHIA, 
KORALZ~\cite{koralz} and KORALW~\cite{koralw} respectively. 
The detector simulation is performed using the 
GEANT~\cite{GEANT} and GHEISHA~\cite{GHEISHA} packages. 
The Monte Carlo
events are reconstructed in the same way as the data. 
Time dependent detector inefficiencies, 
as monitored during the data taking period,
are also simulated.

\section{Event Selection Procedure}

The event selection is performed in two steps. The first one selects
hadronic final states produced in two-photon collisions, the
second identifies a c or b quark by its semileptonic decay.

\subsection{Hadronic Two-Photon Events}
  
Hadronic two-photon events are
selected  by requiring at least five tracks and 
a visible energy, $E_\mathrm{vis}$, below $\sqrt{s} / 3$.
The visible mass, $W_\mathrm{vis}$, 
of the event is calculated from the four-momentum vectors of the
measured particles, tracks and calorimetric clusters
including those from the small angle luminosity monitor. 
These particles
are considered to be pions except for unmatched electromagnetic clusters
considered as photons. $W_\mathrm{vis}$ has to be greater than 3 GeV.
Figures 2a and 2b show the $E_\mathrm{vis}$ and $W_\mathrm{vis}$ 
distributions for the data compared with the Monte Carlo. 
The cut 
on $E_\mathrm{vis}$ separates
the two-photon process from annihilation processes, characterized
by high visible energy.
The
background from $\mathrm{e^{+}e^{-} \ra e^{+}e^{-} \tau^{+} \tau^{-}}$ and
$\mathrm{e^{+}e^{-} \ra \tau^{+} \tau^{-}}$ events is suppressed by
the requirement on the number of tracks.
 
 The analysis is limited to untagged events with small photon virtuality.
Events are excluded when the most energetic cluster in the small
angle calorimeters has an energy 
greater than 0.2 $\mathrm{\sqrt{s}}$. Thus the interacting photons are
quasi-real: $\mathrm{\langle Q^2 \rangle \cong 0.015 \ {GeV}^2}$, 
where $\mathrm{-Q^2}$ is the mass squared of the virtual 
photon.
\subsection{Lepton Selection}

Electrons from charm and beauty semileptonic 
decays are identified by requiring electromagnetic 
clusters in the polar angle range 
$|$cos $\mathrm{\theta| <}$~0.725 with
momentum greater than 0.6 GeV. They should satisfy the following
criteria:

\begin{itemize} 
\item The candidate cluster is required to match to a track.
 The difference between 
  the azimuthal angle between 
the shower barycentre and the track impact point 
  at the calorimeter must be less than 20 mrad.
\item The distribution of energies measured in the crystals 
of the calorimeter should be compatible with that
  of an electromagnetic cluster. 
\item The $\mathrm{E_{t}/p_{t}}$ ratio must be equal to one 
within $\mathrm{2 \sigma}$, 
where $\mathrm{E_{t}}$ is the projection of the energy of the cluster 
   on the plane transverse to the beam, $\mathrm{p_{t}}$ 
is the transverse momentum of the track and $\mathrm{\sigma}$ is
the resolution on this ratio.  

\item Photon conversions are suppressed by requiring 
the distance of closest approach of the track to the mean
 $\mathrm{e^{+} e^{-}}$ collision point in the 
 transverse plane to be less than 0.5 mm and 
the invariant mass of the electron candidate and of the closest
track to be greater than 0.1 GeV.

\end{itemize}

After these cuts, 2434 events remain.
  The background from 
annihilation processes and two-photon production of tau pairs is estimated to be 0.75\%.
  
 Muon candidates are selected from tracks in the muon chambers 
in the angular range $|$cos$~\theta|<0.8$. 
The momentum of the muons must be greater than 2.0 GeV. 
To suppress background from
annihilation processes, the muon momentum must be less
than 0.1 $\mathrm{\sqrt{s}}$.  
After all cuts are applied, 269 events remain.
  The estimated background from 
annihilation processes and two-photon production of tau pairs is 6.0\%. 

\section{Total cross section \boldmath{$\sigma(\mathrm{e^+e^- \rightarrow 
e^+e^- b \bar{b} X})$}}
 
The b cross section is derived from a fit to
the data distributions for events where the momentum 
of the muon or electron 
candidate is greater than 2.0 GeV, corresponding to 269 and 137 events
respectively.
The b selection
efficiency is 2.20\% for muons and 1.25\% for electrons.
The transverse momentum of the lepton with respect to the
nearest jet, $\mathrm{P_{t}}$, is chosen as 
the fit variable since it has the highest sensitivity
to the b fraction.
The jets are reconstructed using the JADE algorithm~\cite{JADE} with 
$\mathrm{y_{cut}=0.1}$. The energy of the muon or electron
is not included in the jet. 

A $\mathrm{\chi^2}$ fit is performed to the data distributions using the sum of the Monte Carlo
distributions of the non two-photon background $N_\mathrm{bkg}$, 
light quark events $N_\mathrm{uds}$, 
c quark events $N_\mathrm{c\bar{c}}$ and
b quark events $N_\mathrm{b\bar{b}}$. 
A three parameter fit is applied in which the
number of beauty, charm and light quarks are free parameters, whereas
$N_\mathrm{bkg}$ is held fixed according to the Monte Carlo prediction.
The results of the fit are given in Table~\ref{tab:fit_result}
which shows also the charm cross section estimate 
from the fit of the muon and electron $\mathrm{P_{t}}$ spectra.

The fit for
muons yields a b
fraction of $51.6 \pm 9.8$ (stat) \%. 
As for the electrons, the b fraction is $42.3 \pm 11.4$ (stat) \%. 
The $\mathrm{\chi^2}$ per degree of freedom  
for the muon and electron fits are $6.2 / 6$ and $10.1 / 6$ respectively.
If no b contribution is included in the fit, confidence levels of
$2.2 \times 10^{-5}$ and $1.2 \times 10^{-3}$ are 
obtained for muon and electron respectively.
The signal 
events are produced in two separate samples for direct and resolved
processes assuming a 1:1 ratio~\cite{theory}.
The fitted distributions are shown 
in Figures~\ref{fig:ptmu_fit} and \ref{fig:pte_fit}. 

The resulting cross sections for the luminosity averaged centre-of-mass
energy
$\langle \sqrt{s} \rangle = 194$ GeV are:
    \[ \mathrm{\sigma(e^+e^-\rightarrow e^+e^-b \bar b X)}_{\mbox{muons}}=  
 14.9 \pm{2.8}\ \mbox{(stat)}\ \pm{2.6}\ \mbox{(syst)}\  
                \mbox{pb} \]
  \[ \mathrm{\sigma(e^+e^-\rightarrow e^+e^-b \bar b X)}_{\mbox{electrons}}=  
 10.9 \pm{2.9}\ \mbox{(stat)}\ \pm{2.0}\ \mbox{(syst)}\  
                \mbox{pb}. \]
The systematic
uncertainties arise from the event selection, jet reconstruction,
massive or massless charm quarks in the event generation,
b semileptonic
branching ratio, 
trigger efficiency, Monte Carlo statistics and 
direct to resolved process ratio. Table \ref{tab:syst_bb} shows the
values of systematic uncertainties due to these contributions.
The dominant uncertainty comes from 
the event selection. It is estimated
by variation of the cuts and includes detector resolution uncertainties
and the agreement between data and Monte Carlo.
The systematic uncertainty on jet reconstruction
is assigned by variation of the $\mathrm{y_{cut}}$ parameter. 
The contribution due to the uncertainty on the ratio of the direct to
the resolved processes is estimated by changing it
from 1:1 to 1:2 or 2:1. 

The combination of muon and electron results gives
    \[ \mathrm{\sigma(e^+e^-\rightarrow e^+e^-b \bar b X)}_{\mbox{combined}}=  
 13.1 \pm{2.0}\ \mbox{(stat)}\ \pm{2.4}\ \mbox{(syst)}\  
                \mbox{pb}. \]

\section{Total cross section 
\boldmath{$\sigma(\mathrm{e^+e^- \rightarrow e^+e^- c \bar{c} X})$}}

In order to increase the statistical accuracy for the charm cross 
section measurement, the electron momentum cut 
is relaxed from 2 GeV to 0.6 GeV. 
The cross section is
calculated as for the data collected at centre-of mass energies from
91 GeV to 183 GeV~\cite{L3}.
In addition, the beauty contribution to the number of observed events
is subtracted.
The open charm cross section is then
calculated from the number of events with leptons 
using the equation:
\begin{equation}
  \mathrm{\sigma(e^+e^-\rightarrow e^+e^-c \bar c X)} = \frac{(N_{\mathrm{obs}}^{\mathrm{lept}}-N_{\mathrm{bkg}}^{\mathrm{lept}}-N_{\mathrm{b}}^{\mathrm{lept}})~\pi_{\mathrm{c}}}{{\cal{L}}~\epsilon_{\mathrm{trig}}~\epsilon_\mathrm{c}^{\prime}},
\label{eq:cross}
\end{equation}
where the variables are defined as follows:
\begin{itemize}
 
\item $N_{\mathrm{obs}}^{\mathrm{lept}}$ is the number of events 
in the data after the final electron selection (2434 events).  
  
\item $\epsilon_{\mathrm{trig}}$ is the trigger efficiency (94.4\%)
which is determined 
from the data using a set of independent triggers.    

\item $N_{\mathrm{bkg}}^{\mathrm{lept}}$ is the number of 
background events which do not
  originate from two-photon hadronic interactions 
estimated from Monte Carlo ($18.3 \pm 2.0$ events).

\item $N_{\mathrm{b}}^{\mathrm{lept}}$ is the number of 
beauty events (169.5) estimated 
from the cross section measured above.

\item $\cal{L}$ is the total integrated luminosity (410 $\mathrm{pb^{-1}}$). 

\end{itemize}
 The c selection efficiency, $\mathrm{\epsilon_c^{\prime}}$, is the 
  fraction of c events selected
relative to those generated in the full phase space.  
In order to be
independent of the Monte Carlo flavour composition, 
the charm purity is written as:
  \begin{equation}
    \pi_\mathrm{c} = (1-\frac{\epsilon_{\mathrm{uds}}}{\epsilon_{\mathrm{data}}}) / 
            (1-\frac{\epsilon_{\mathrm{uds}}}{\epsilon_{\mathrm{c}}}),
    \label{eq:purityeff}
  \end{equation}
where $\mathrm{\epsilon_{c}}$ and $\mathrm{\epsilon_{uds}}$ are the
  fractions of charm events, $N_\mathrm{c}^{\mathrm{lept}}$, and 
light quark events, $N_{\mathrm{uds}}^{\mathrm{lept}}$, 
  accepted by the final selection. 
Equation (2) follows from expressing the
number of non-beauty hadronic events as:
  \begin{equation}
    \frac{N_\mathrm{c}^{\mathrm{lept}} + N_{\mathrm{uds}}^{\mathrm{lept}}}{\epsilon_{\mathrm{data}}} = \frac{N_\mathrm{c}^{\mathrm{lept}}}{\epsilon_\mathrm{c}} + 
\frac{N_{\mathrm{uds}}^{\mathrm{lept}}}{\epsilon_{\mathrm{uds}}},
    \label{eq:purityderive}
  \end{equation}
where the quantity $\mathrm{\epsilon_{data}}$ is 
defined by the following relation:  
  \begin{equation}
 \epsilon_{\mathrm{data}} = \frac{N_\mathrm{c}^{\mathrm{lept}}+N_{\mathrm{uds}}^{\mathrm{lept}}}{N_\mathrm{c}^{\mathrm{had}}+N_{\mathrm{uds}}^{\mathrm{had}}}= \frac{N_{\mathrm{obs}}^{\mathrm{lept}}-N_{\mathrm{bkg}}^{\mathrm{lept}}-N_{\mathrm{b}}^{\mathrm{lept}}}{N_{\mathrm{obs}}^{\mathrm{had}}-N_{\mathrm{bkg}}^{\mathrm{had}}-N_{\mathrm{b}}^{\mathrm{had}}}.
    \label{eq:epsgg}
  \end{equation}
$N_\mathrm{b}^{\mathrm{had}}$, $N_\mathrm{c}^{\mathrm{had}}$ 
and $N_\mathrm{uds}^{\mathrm{had}}$ are the number of hadronic
events with b-quarks, c-quarks and light quarks respectively.
$N_\mathrm{obs}^{\mathrm{had}}$ and $N_\mathrm{bkg}^{\mathrm{had}}$
are the observed number of hadronic events and the background
expectations respectively.
This method is 
insensitive to the absolute normalization of the c and uds background
Monte Carlo samples, 
but still depends on the ratio of direct to resolved processes
in the signal Monte Carlo.
 
For the electron sample, the charm purity is 75.0\% 
 and the charm selection 
efficiency is 0.41\%.
The charm production cross section 
in $\mathrm{\gamma \gamma}$ collisions 
at $\langle \sqrt{s} \rangle = 194$ GeV is then:
  \begin{equation}
    \mathrm{\sigma(e^+e^-\rightarrow e^+e^-c \bar c X)}_{\mathrm{electrons}}=  
 1072 \pm{33}\ \mathrm{(stat)}\ \pm{126}\ \mathrm{(syst)}\  
                \mathrm{pb}.
    \label{eq:charm}
  \end{equation}
This charm cross section is compatible with the fit results
reported in Table~\ref{tab:fit_result}. 
The systematic
uncertainties arise from the event selection, direct to resolved process ratio,
c semileptonic branching ratio, massive or massless charm quarks in the event generation,
 experimental uncertainties on the beauty cross section,
trigger efficiency, Monte Carlo statistics and uds background estimate. 
 The average charm semileptonic branching ratio used in the 
simulation is 0.098~\cite{branching}.
Table~\ref{tab:syst_cc} shows the
values of the systematic uncertainties due to different contributions.
The dominant systematic
uncertainty is from event selection 
and from the variation of the ratio of direct to 
resolved processes.
They are estimated as in the beauty study.

The fit result for beauty production is checked by this counting 
method in the electron case, fixing the charm cross section 
to the value of equation (5).
In addition to the momentum
cut of 2 GeV, the transverse momentum $\mathrm{P_{t}}$ 
is required to be greater than 1.0 GeV. 
After all cuts are applied 106 electron candidates remain.
The beauty purity is 49.0\%, and the selection efficiency
is 1.2\%.
The cross section at 
$\langle \sqrt{s} \rangle = 194$ GeV is
    \[ \mathrm{\sigma(e^+e^-\rightarrow e^+e^-b \bar b X)}_{\mbox{electrons}}=  
 11.3 \pm{2.3}\ \mbox{(stat)}\ \mbox{pb}, \]
in good agreement with the fit result.

Figure~\ref{fig:pte_cut} shows the momentum distribution 
of the electron candidates.
In this plot the charm and beauty 
cross sections predicted by 
the PYTHIA Monte Carlo are scaled to the measured values.
The data and Monte Carlo shapes of the momentum 
distribution show a good agreement.

In the case of muons, where the 2 GeV cut can not
be relaxed, the estimation of charm 
production is that derived by a 
simultaneous fit to the b and c fractions
described in the previous section:
    \[ \mathrm{\sigma(e^+e^-\rightarrow e^+e^-c \bar c X)}_{\mbox{muons}}=  
 814 \pm{164}\ \mbox{(stat)}\ \pm{200}\ \mbox{(syst)}\  
                \mbox{pb} \]
at $\langle \sqrt{s} \rangle = 194$ GeV.
The efficiency is much lower for the muon sample, about 0.04\%, 
due to the higher momentum cut. 
The dominant systematic
uncertainties are the event selection  
and the direct to resolved process ratio. 

The combined value for the open charm cross section at 
$\langle \sqrt{s} \rangle = 194$ GeV is:   
    \[ \mathrm{\sigma(e^+e^-\rightarrow e^+e^-c \bar c X)}_{\mbox{combined}}=  
 1016 \pm{30}\ \mbox{(stat)}\ \pm{120}\ \mbox{(syst)}\  
                \mbox{pb}. \]

\section{Comparisons with QCD Predictions}

The cross sections 
for open beauty and charm production are compared in Figure 6 
to perturbative next-to-leading order QCD calculations~\cite{theory}.
 The dashed line corresponds to the direct process,
NLO QCD calculations, while the solid line shows the prediction
for the sum of direct and resolved processes.
The direct process 
depends upon the QCD coupling constant and the heavy-quark mass.  
The prediction for open charm is calculated
using a charm mass of either 1.3 GeV or 1.7 GeV and the open charm 
threshold energy is set to 3.8 GeV.
The theory prediction for the resolved process is calculated
with the GRV parton density function~\cite{GRV}.
The
renormalization and factorization scales are chosen to be the heavy quark mass. 
The direct process $\mathrm{\gamma \gamma \rightarrow c
\bar{c}}$~ is insufficient to describe the data, even if real and 
virtual gluon corrections are included. The data 
therefore require a significant gluon content in the photon.

The prediction for open beauty is calculated
for a b quark mass of 4.5 GeV or 5.0 GeV and the open beauty 
threshold energy is set to 10.6 GeV.
For $\langle \sqrt{s} \rangle = 194$ GeV and a b quark mass of 4.5 GeV,
this cross section is 4.4 pb.
The $\mathrm{b \bar{b}}$ cross section
is measured in $\mathrm{\gamma\gamma}$ collisions for the first time
and is a factor of 3 and about 4 statistical uncertainty standard 
deviations higher than expected.

\section*{Acknowledgements}

We wish to express our gratitude to the CERN accelerator divisions for the 
excellent performance of the LEP machine. We also acknowledge 
and appreciate 
the effort of the engineers, technicians and support staff who have
participated in the construction and maintenance of this experiment.

%

\clearpage
\newpage
%
%
\section*{Author List}
\typeout{   }     
\typeout{Using author list for paper 226 -- ? }
\typeout{$Modified: Tue Sep  5 19:04:46 2000 by smele $}
\typeout{!!!!  This should only be used with document option a4p!!!!}
\typeout{   }
%
%
%
%
%
%

\newcount\tutecount  \tutecount=0
\def\tutenum#1{\global\advance\tutecount by 1 \xdef#1{\the\tutecount}}
\def\tute#1{$^{#1}$}
\tutenum\aachen            
\tutenum\nikhef            
\tutenum\mich              
\tutenum\lapp              
\tutenum\basel             
\tutenum\lsu               
\tutenum\beijing           
\tutenum\berlin            
\tutenum\bologna           
\tutenum\tata              
\tutenum\ne                
\tutenum\bucharest         
\tutenum\budapest          
\tutenum\mit               
\tutenum\debrecen          
\tutenum\florence          
\tutenum\cern              
\tutenum\wl                
\tutenum\geneva            
\tutenum\hefei             
\tutenum\seft              
\tutenum\lausanne          
\tutenum\lecce             
\tutenum\lyon              
\tutenum\madrid            
\tutenum\milan             
\tutenum\moscow            
\tutenum\naples            
\tutenum\cyprus            
\tutenum\nymegen           
\tutenum\caltech           
\tutenum\perugia           
\tutenum\cmu               
\tutenum\prince            
\tutenum\rome              
\tutenum\peters            
\tutenum\potenza           
\tutenum\riverside         
\tutenum\salerno           
\tutenum\ucsd              
\tutenum\santiago          
\tutenum\sofia             
\tutenum\korea             
\tutenum\alabama           
\tutenum\utrecht           
\tutenum\purdue            
\tutenum\psinst            
\tutenum\zeuthen           
\tutenum\eth               
\tutenum\hamburg           
\tutenum\taiwan            
\tutenum\tsinghua          

{
\parskip=0pt
\noindent
{\bf The L3 Collaboration:}
\ifx\selectfont\undefined
 \baselineskip=10.8pt
 \baselineskip\baselinestretch\baselineskip
 \normalbaselineskip\baselineskip
 \ixpt
\else
 \fontsize{9}{10.8pt}\selectfont
\fi
\medskip
\tolerance=10000
\hbadness=5000
\raggedright
\hsize=162truemm\hoffset=0mm
\def\r{\rlap,}
\noindent

M.Acciarri\r\tute\milan\
P.Achard\r\tute\geneva\ 
O.Adriani\r\tute{\florence}\ 
M.Aguilar-Benitez\r\tute\madrid\ 
J.Alcaraz\r\tute\madrid\ 
G.Alemanni\r\tute\lausanne\
J.Allaby\r\tute\cern\
A.Aloisio\r\tute\naples\ 
M.G.Alviggi\r\tute\naples\
G.Ambrosi\r\tute\geneva\
% ...
\newpage
\begin{list}{A}{\itemsep=0pt plus 0pt minus 0pt\parsep=0pt plus 0pt minus 0pt
                \topsep=0pt plus 0pt minus 0pt}
\item[\aachen]
 I. Physikalisches Institut, RWTH, D-52056 Aachen, FRG$^{\S}$\\
 III. Physikalisches Institut, RWTH, D-52056 Aachen, FRG$^{\S}$
\item[\nikhef] National Institute for High Energy Physics, NIKHEF, 
     and University of Amsterdam, NL-1009 DB Amsterdam, The Netherlands
\item[\mich] University of Michigan, Ann Arbor, MI 48109, USA
\item[\lapp] Laboratoire d'Annecy-le-Vieux de Physique des Particules, 
     LAPP,IN2P3-CNRS, BP 110, F-74941 Annecy-le-Vieux CEDEX, France
\item[\basel] Institute of Physics, University of Basel, CH-4056 Basel,
     Switzerland
\item[\lsu] Louisiana State University, Baton Rouge, LA 70803, USA
\item[\beijing] Institute of High Energy Physics, IHEP, 
  100039 Beijing, China$^{\triangle}$ 
\item[\berlin] Humboldt University, D-10099 Berlin, FRG$^{\S}$
\item[\bologna] University of Bologna and INFN-Sezione di Bologna, 
     I-40126 Bologna, Italy
\item[\tata] Tata Institute of Fundamental Research, Bombay 400 005, India
\item[\ne] Northeastern University, Boston, MA 02115, USA
\item[\bucharest] Institute of Atomic Physics and University of Bucharest,
     R-76900 Bucharest, Romania
\item[\budapest] Central Research Institute for Physics of the 
     Hungarian Academy of Sciences, H-1525 Budapest 114, Hungary$^{\ddag}$
\item[\mit] Massachusetts Institute of Technology, Cambridge, MA 02139, USA
\item[\debrecen] KLTE-ATOMKI, H-4010 Debrecen, Hungary$^\P$
\item[\florence] INFN Sezione di Firenze and University of Florence, 
     I-50125 Florence, Italy
\item[\cern] European Laboratory for Particle Physics, CERN, 
     CH-1211 Geneva 23, Switzerland
\item[\wl] World Laboratory, FBLJA  Project, CH-1211 Geneva 23, Switzerland
\item[\geneva] University of Geneva, CH-1211 Geneva 4, Switzerland
\item[\hefei] Chinese University of Science and Technology, USTC,
      Hefei, Anhui 230 029, China$^{\triangle}$
\item[\lausanne] University of Lausanne, CH-1015 Lausanne, Switzerland
\item[\lecce] INFN-Sezione di Lecce and Universit\`a Degli Studi di Lecce,
     I-73100 Lecce, Italy
\item[\lyon] Institut de Physique Nucl\'eaire de Lyon, 
     IN2P3-CNRS,Universit\'e Claude Bernard, 
     F-69622 Villeurbanne, France
\item[\madrid] Centro de Investigaciones Energ{\'e}ticas, 
     Medioambientales y Tecnolog{\'\i}cas, CIEMAT, E-28040 Madrid,
     Spain${\flat}$ 
\item[\milan] INFN-Sezione di Milano, I-20133 Milan, Italy
\item[\moscow] Institute of Theoretical and Experimental Physics, ITEP, 
     Moscow, Russia
\item[\naples] INFN-Sezione di Napoli and University of Naples, 
     I-80125 Naples, Italy
\item[\cyprus] Department of Natural Sciences, University of Cyprus,
     Nicosia, Cyprus
\item[\nymegen] University of Nijmegen and NIKHEF, 
     NL-6525 ED Nijmegen, The Netherlands
\item[\caltech] California Institute of Technology, Pasadena, CA 91125, USA
\item[\perugia] INFN-Sezione di Perugia and Universit\`a Degli 
     Studi di Perugia, I-06100 Perugia, Italy   
\item[\cmu] Carnegie Mellon University, Pittsburgh, PA 15213, USA
\item[\prince] Princeton University, Princeton, NJ 08544, USA
\item[\rome] INFN-Sezione di Roma and University of Rome, ``La Sapienza",
     I-00185 Rome, Italy
\item[\peters] Nuclear Physics Institute, St. Petersburg, Russia
\item[\potenza] INFN-Sezione di Napoli and University of Potenza, 
     I-85100 Potenza, Italy
\item[\riverside] University of Californa, Riverside, CA 92521, USA
\item[\salerno] University and INFN, Salerno, I-84100 Salerno, Italy
\item[\ucsd] University of California, San Diego, CA 92093, USA
\item[\santiago] Dept. de Fisica de Particulas Elementales, Univ. de Santiago,
     E-15706 Santiago de Compostela, Spain
\item[\sofia] Bulgarian Academy of Sciences, Central Lab.~of 
     Mechatronics and Instrumentation, BU-1113 Sofia, Bulgaria
\item[\korea]  Laboratory of High Energy Physics, 
     Kyungpook National University, 702-701 Taegu, Republic of Korea
\item[\alabama] University of Alabama, Tuscaloosa, AL 35486, USA
\item[\utrecht] Utrecht University and NIKHEF, NL-3584 CB Utrecht, 
     The Netherlands
\item[\purdue] Purdue University, West Lafayette, IN 47907, USA
\item[\psinst] Paul Scherrer Institut, PSI, CH-5232 Villigen, Switzerland
\item[\zeuthen] DESY, D-15738 Zeuthen, 
     FRG
\item[\eth] Eidgen\"ossische Technische Hochschule, ETH Z\"urich,
     CH-8093 Z\"urich, Switzerland
\item[\hamburg] University of Hamburg, D-22761 Hamburg, FRG
\item[\taiwan] National Central University, Chung-Li, Taiwan, China
\item[\tsinghua] Department of Physics, National Tsing Hua University,
      Taiwan, China
\item[\S]  Supported by the German Bundesministerium 
        f\"ur Bildung, Wissenschaft, Forschung und Technologie
\item[\ddag] Supported by the Hungarian OTKA fund under contract
numbers T019181, F023259 and T024011.
\item[\P] Also supported by the Hungarian OTKA fund under contract
  numbers T22238 and T026178.
\item[$\flat$] Supported also by the Comisi\'on Interministerial de Ciencia y 
        Tecnolog{\'\i}a.
\item[$\sharp$] Also supported by CONICET and Universidad Nacional de La Plata,
        CC 67, 1900 La Plata, Argentina.
\item[$\diamondsuit$] Also supported by Panjab University, Chandigarh-160014, 
        India.
\item[$\triangle$] Supported by the National Natural Science
  Foundation of China.
\end{list}
}
\vfill


\newpage
\clearpage
\begin{table}[htb]
  \vspace{0.2cm}
  \begin{center}
    \begin{tabular}{|c||c|c|} \hline
 Fit results       & Muon Tag & Electron Tag  \\ \hline
\phantom{0}$\mathrm{N_{bkg}}$             & \phantom{0}16.2 (fixed) & \phantom{0}2.9 (fixed) \\ 
$\mathrm{N_{b\bar{b}}}$        & $126.7 \pm 24.1$   & $52.5 \pm 14.1$  \\
$\mathrm{N_{c\bar{c}}}$        & $119.0 \pm 24.0$   & $71.5 \pm 14.8$ \\ 
\phantom{0}$\mathrm{N_{uds}}$             & $0.0^{ + 33.0}_{ - \phantom{0}0.0}$  & $0.0^{ + 7.4}_{ - 0.0}$ \\ 
$\mathrm{\chi^2}$ / d.o.f.     & 6.2 / 6   & 10.1 / 6 \\  \hline 
\rule{0pt}{12pt} $\sigma(\mathrm{e^+e^- \rightarrow e^+e^- b\bar{b} X})$, (pb) & $14.9 \pm 2.8 $   & $10.9 \pm 2.9 $ \\ 
\rule{0pt}{12pt} $\sigma(\mathrm{e^+e^- \rightarrow e^+e^- c\bar{c} X})$, (pb) & $814 \pm 164 $   & $1092 \pm 226 $ \\ \hline 
    \end{tabular}
  \end{center}
  \caption{Fit to the distribution of 
the transverse momentum of the lepton with respect to the
nearest jet. The fit parameters are constrained to be positive. The correlation between $\mathrm{N_{b\bar{b}}}$  and $\mathrm{N_{c\bar{c}}}$ is 75\%.}
  \label{tab:fit_result} 
\end{table}

\begin{table}[htb]
  \vspace{0.2cm}
  \begin{center}
    \begin{tabular}{|l||c|c|} \hline
  Source of uncertainty       & Muon Tag & Electron Tag  \\ \cline{2-3}
      & \rule{0pt}{12pt} $\mathrm{\Delta\sigma(e^{+}e^{-} \ra e^{+}e^{-}b\bar{b}X)}$, \%  & \rule{0pt}{12pt} $\mathrm{\Delta\sigma(e^{+}e^{-} \ra e^{+}e^{-}b\bar{b}X)}$, \% \\ \hline  
Event selection                    & 14.6 & 15.8 \\ 
Jet reconstruction                 & \phantom{0}8.2   & \phantom{0}8.2 \\
Massive/massless charm             & \phantom{0}3.0   & \phantom{0}3.0 \\
$\mathrm{B(b \rightarrow e,\mu)}$  & \phantom{0}2.0   & \phantom{0}2.0 \\
Trigger efficiency                 & \phantom{0}2.0   & \phantom{0}2.0 \\ 
Monte Carlo statistics             & \phantom{0}1.4  & \phantom{0}1.8 \\ 
Direct / resolved ratio            & \phantom{0}1.0   & \phantom{0}0.9 \\ \hline
 Total                             & 17.3   & 18.4  \\ \hline
    \end{tabular}
  \end{center}
  \caption{Systematic uncertainties on $\mathrm{\sigma(e^+e^-\rightarrow e^+e^-b \bar b X)}$.}
  \label{tab:syst_bb} 
\end{table}

\begin{table}[htb]
  \vspace{0.2cm}
  \begin{center}
    \begin{tabular}{|l||c|c|} \hline
  Source of uncertainty       & Electron Tag & Muon Tag  \\ \cline{2-3}
      & $\mathrm{\Delta\sigma(e^{+}e^{-} \ra e^{+}e^{-}c\bar{c}X)}$, \%  & $\mathrm{\Delta\sigma(e^{+}e^{-} \ra e^{+}e^{-}c\bar{c}X)}$, \% \\ \hline
Event selection                    & \phantom{0}8.5   & 18.6 \\
Direct / resolved ratio            & \phantom{0}5.7   & 10.9 \\  
$\mathrm{B(c \rightarrow e,\mu)}$  & \phantom{0}3.0   & \phantom{0}5.0 \\ 
Massive/massless charm             & \phantom{0}3.0   & \phantom{0}3.0 \\
b background                       & \phantom{0}2.4   & \phantom{0}---  \\
Trigger efficiency                 & \phantom{0}2.0   & \phantom{0}2.0 \\ 
Monte Carlo statistics             & \phantom{0}1.9  & \phantom{0}6.0 \\ 
uds background                     & \phantom{0}1.1   & \phantom{0}---  \\ 
Jet reconstruction                 & \phantom{0}---   & \phantom{0}8.2 \\ \hline
 Total                             & 11.7   & 24.6  \\ \hline
    \end{tabular}
  \end{center}
  \caption{Systematic uncertainties on $\mathrm{\sigma(e^+e^-\rightarrow e^+e^-c \bar c X)}$.}
  \label{tab:syst_cc} 
\end{table}

\newpage
\clearpage

\newpage

\begin{figure}[htbp]
  \begin{tabular}{cc}
    \mbox{\epsfig{file=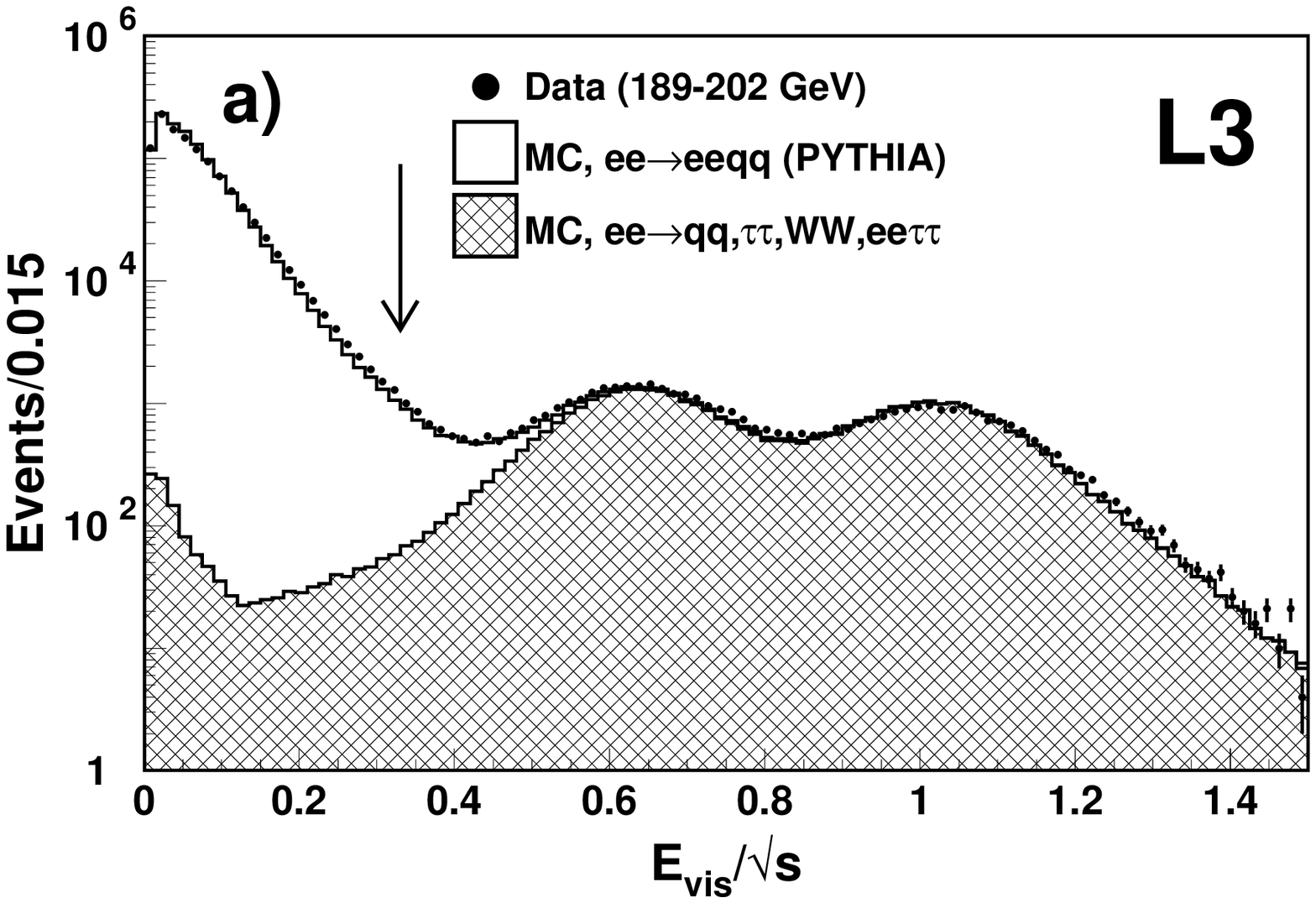, width=.9\textwidth}} \\
    \mbox{\epsfig{file=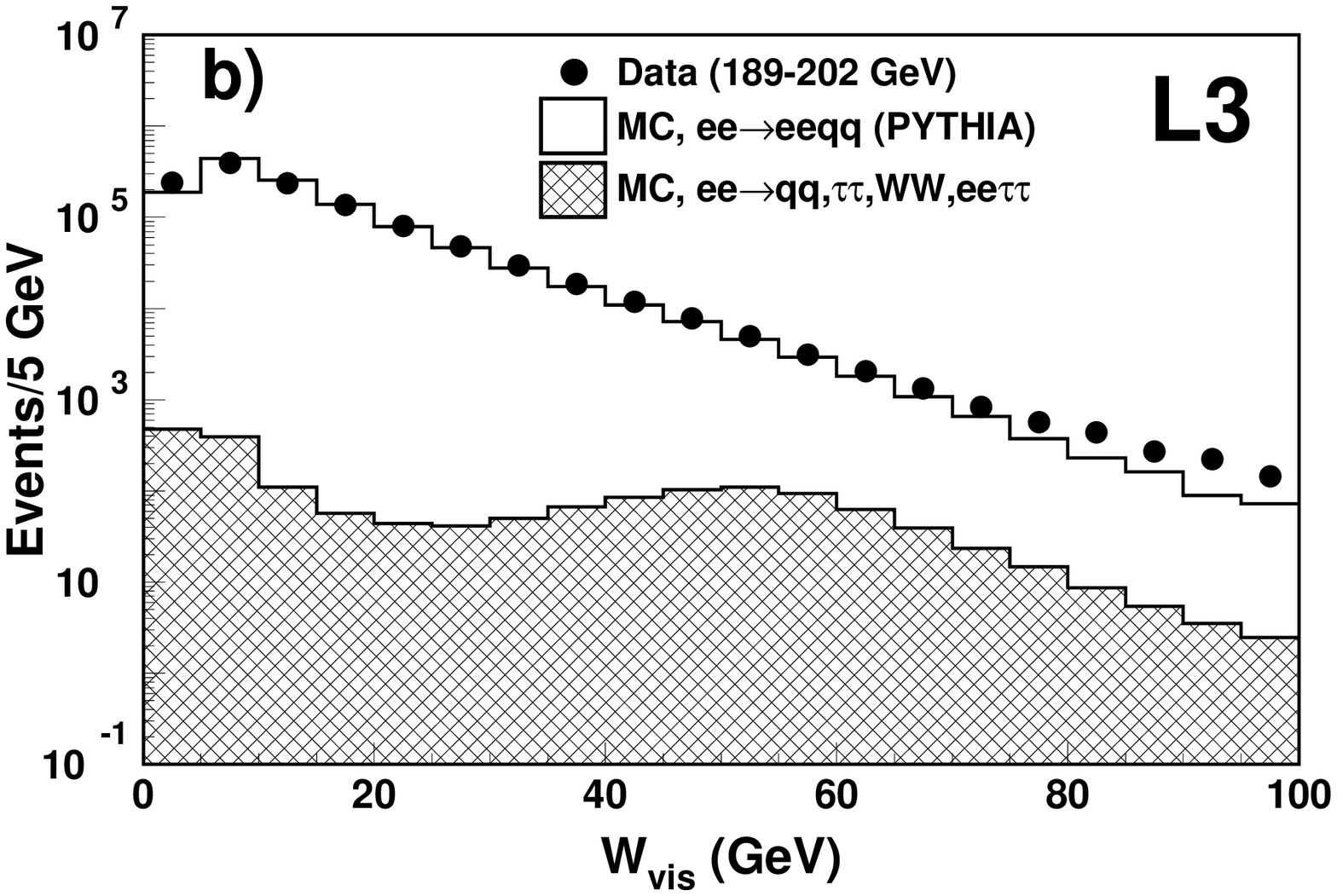, width=.9\textwidth}} 
  \end{tabular}
  \caption{a) Total visible energy after applying 
hadronic selection cuts. A cut of $E_\mathrm{vis} < 0.33\sqrt{s}$ 
indicated by the arrow
removes most of the backgrounds. 
b) The visible mass after applying hadronic selection cuts.}
 \label{fig:eviswvis}
\end{figure}

\newpage

\begin{figure}[htbp]
  \begin{tabular}{c}
    \mbox{\epsfig{file=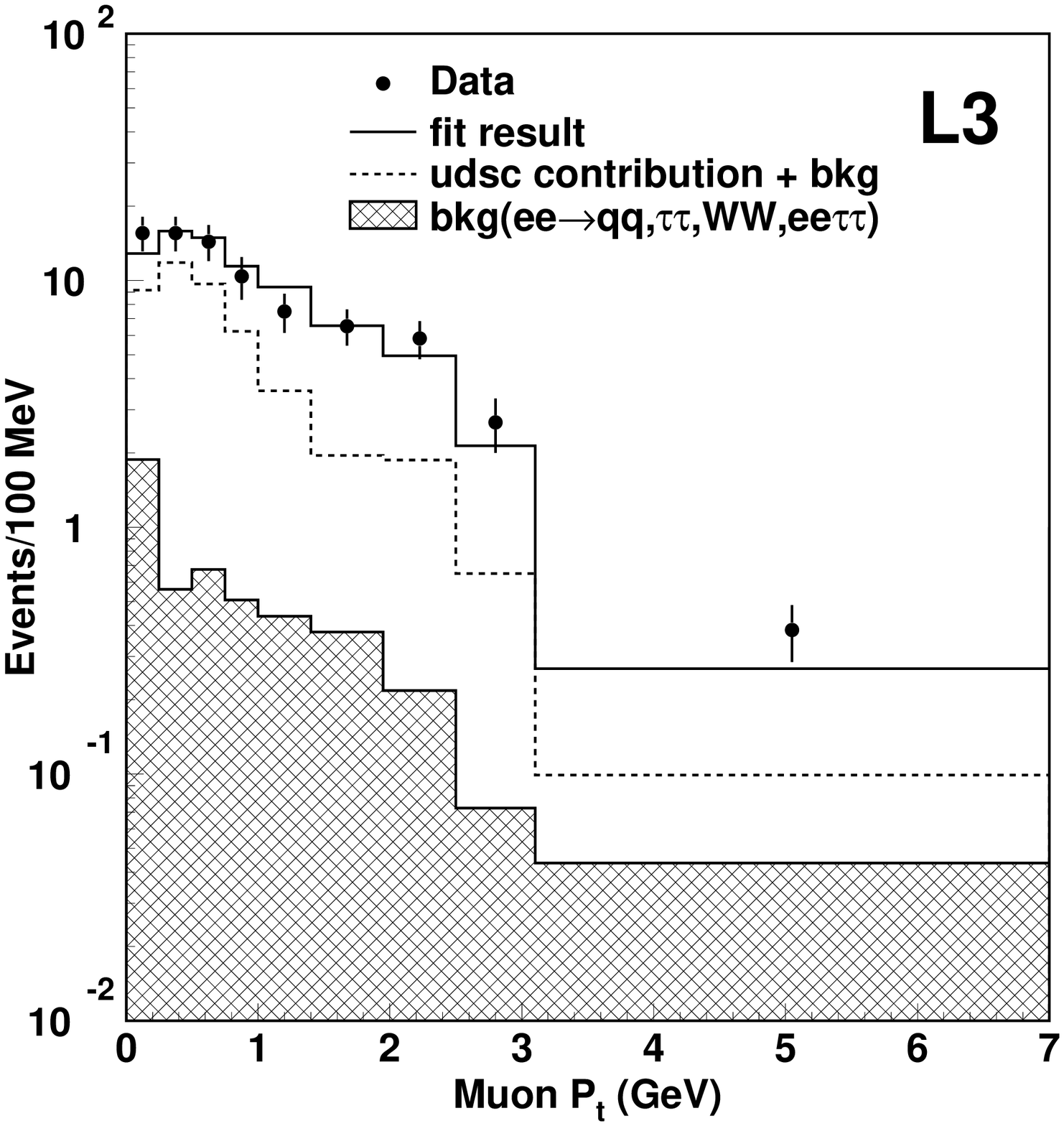, width=.9\textwidth}} \\
  \end{tabular}
  \caption{The distribution of the transverse momentum, $\mathrm{P_{t}}$,
of the muon candidate with respect to the closest jet.}
  \label{fig:ptmu_fit}
\end{figure}

\newpage

\begin{figure}[htbp]
  \begin{tabular}{c}
    \mbox{\epsfig{file=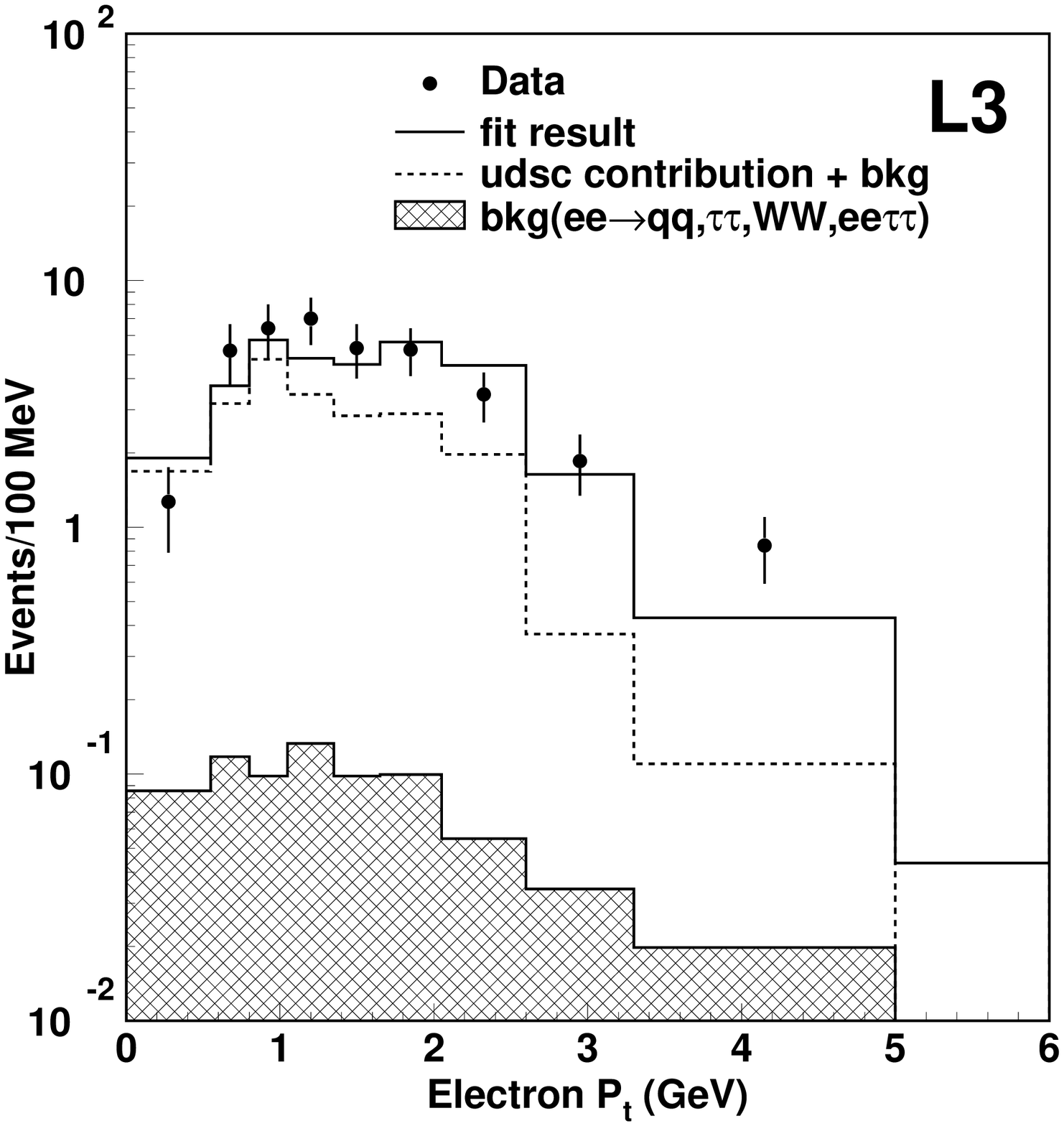, width=.9\textwidth}} \\
  \end{tabular}
  \caption{The distribution of the transverse momentum, $\mathrm{P_{t}}$,
of the electron candidate with respect to the closest jet.}
  \label{fig:pte_fit}
\end{figure}

\newpage

\begin{figure}[htbp]
  \begin{tabular}{c}
    \mbox{\epsfig{file=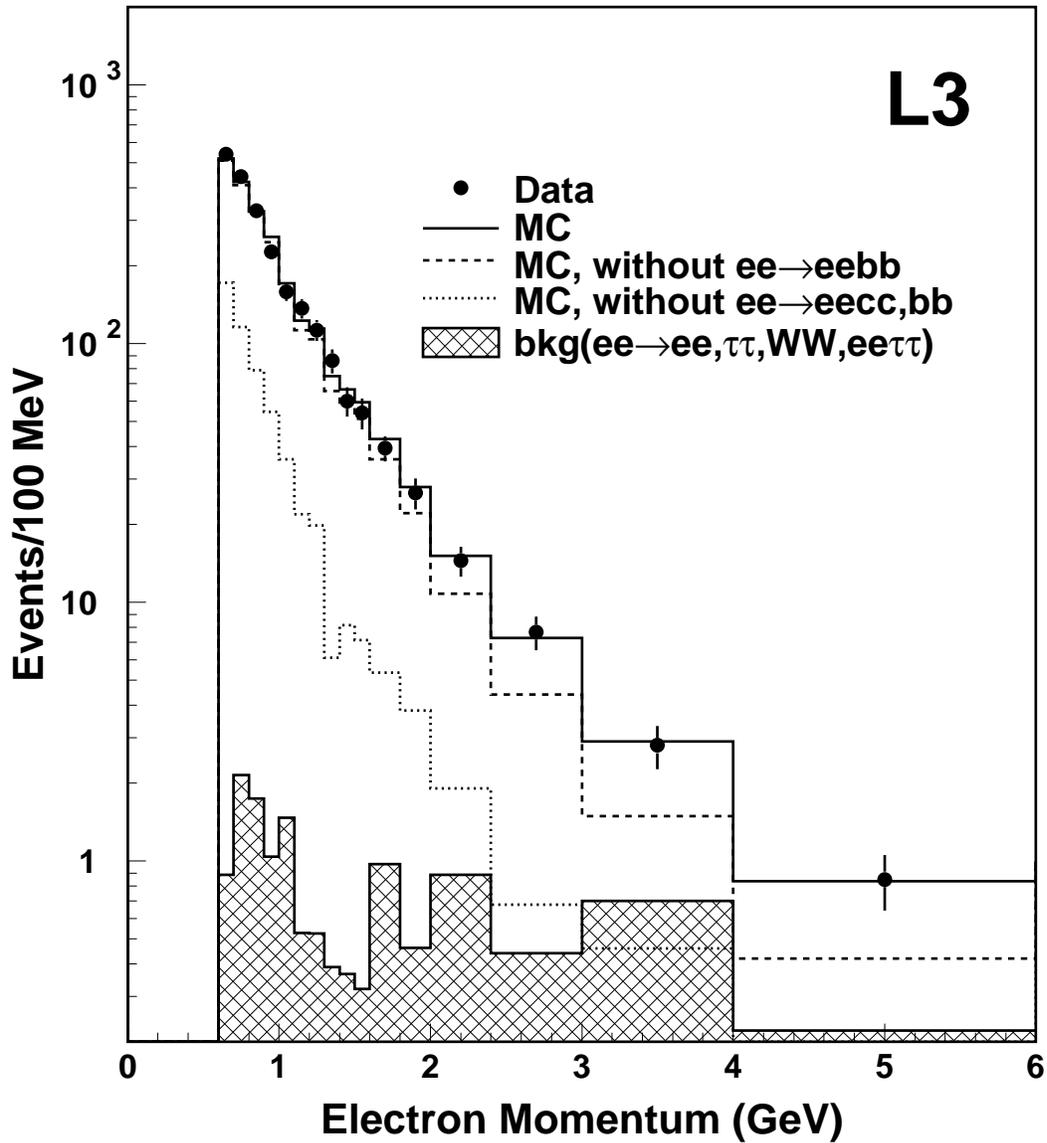, width=.9\textwidth}} \\
  \end{tabular}
  \caption{The momentum of the electron
candidates. The dotted, dashed and solid 
histograms are the contributions 
of uds, udsc and udscb quarks from the PYTHIA Monte Carlo.
The c and b fraction of PYTHIA are scaled to the measured cross sections.}
  \label{fig:pte_cut}
\end{figure}

\newpage

\begin{figure}[htbp]
 \mbox{\epsfig{file=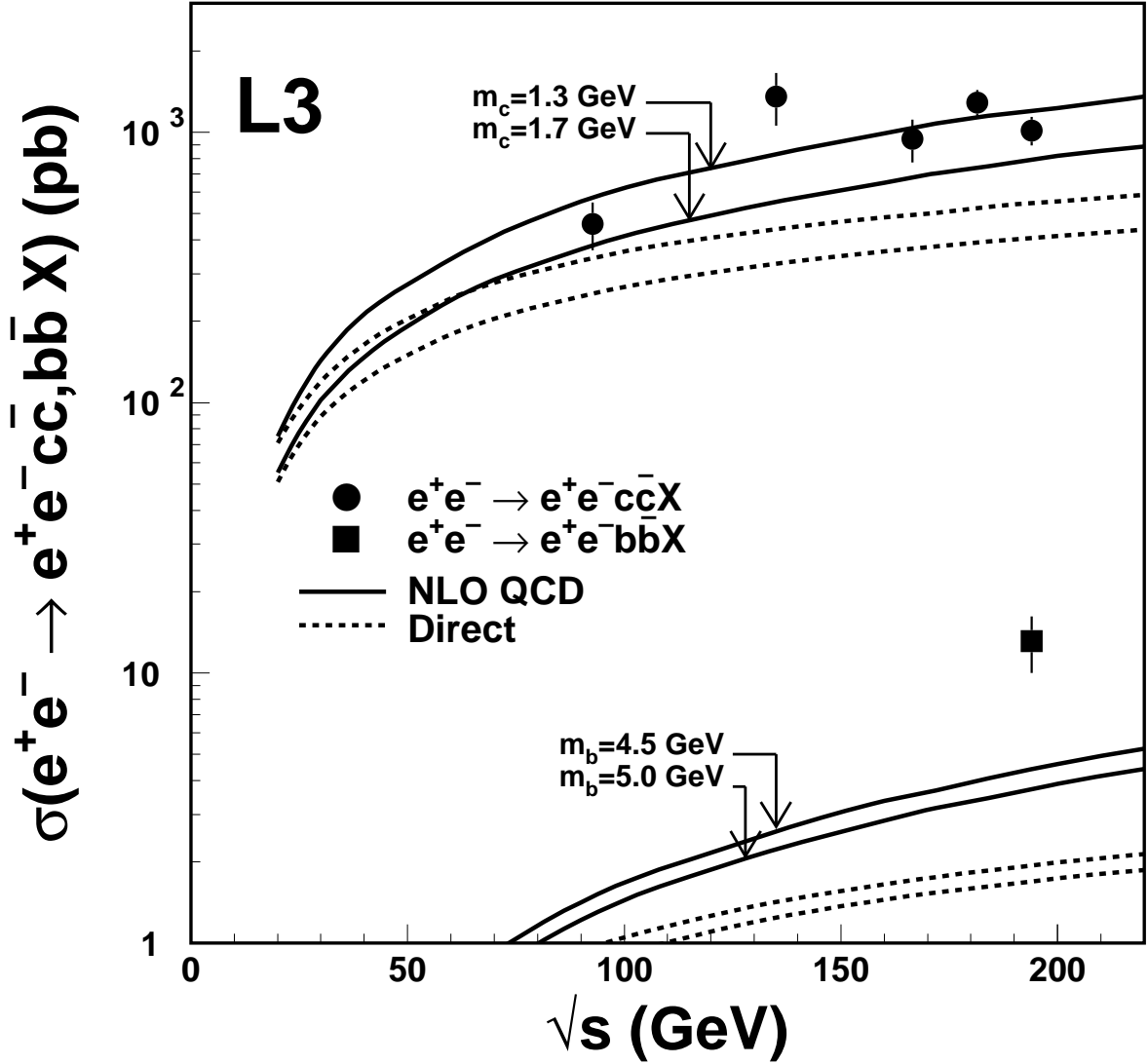,width=17cm}}
    \caption{The open charm and beauty production cross section 
      in two-photon collisions. The
      L3 data from both electron and muon events 
      are combined. The statistical and systematic uncertainties
      are added in quadrature. The dashed line corresponds
      to the direct process contribution and the solid
      line represents the NLO QCD prediction for the sum of
      the direct and resolved processes.}
    \label{fig:sigma_ccbb}
\end{figure}

\end{document}